\documentclass[twocolumn,showpacs,preprintnumbers,amsmath,amssymb,prl]{revtex4}

\usepackage{graphicx}
\usepackage{dcolumn}
\usepackage{epsf}

\begin{document}
\title{Reduction of Dissipative Nonlinear Conductivity of Superconductors by Static and Microwave Magnetic Fields.}
\author{A. Gurevich\email{gurevich@odu.edu}}
\affiliation{Department of Physics and Center for Accelerator Science, Old Dominion University, 4600 Elkhorn Avenue
Norfolk, Virginia 23529, USA}
\begin{abstract}
A theory of dissipative nonlinear conductivity, $\sigma_1(\omega,H)$, of s-wave superconductors under strong electromagnetic fields 
at low temperatures is proposed. Closed-form expressions for $\sigma_1(H)$ and the 
surface resistance $R_s(\omega,H)$ are obtained in the nonequilibrium dirty limit for which $\sigma_1(H)$ has a significant minimum 
as a function of a low-frequency $(\hbar\omega\ll k_BT)$ magnetic field $H$.  The calculated microwave suppression of $R_s(H)$ is in good agreement with 
recent experiments on alloyed Nb resonator cavities. It is shown that superimposed dc and ac fields, $H=H_0+H_a\cos\omega t$,  
can be used to reduce ac dissipation in thin film nanostructures by tuning $\sigma_1(H_0)$ with the dc field.    

\end{abstract}


\maketitle

One of the hallmarks of superconductivity is that static magnetic fields $H$ induce screening currents that break Cooper pairs and 
reduce the transition temperature $T_c$ \cite{Bardeen}. This manifests itself in the nonlinear Meissner 
effect \cite{nme} and intermodulation \cite{Dahm}, which have been observed on high-$T_c$ cuprates \cite{hein,oates}. Behavior 
of a superconductor becomes far more complex under the alternating field $H=H_a\cos\omega t$, which not only induces 
pairbreaking currents, but also drives the quasiparticles out of equilibrium, particularly if the frequency $\omega$ exceeds the superconducting gap $\Delta$ \cite{kopnin}. 
Microwave absorption can produce nonequilibrium states with higher $T_c$ and the critical current $I_c$ as has been observed on thin films and tunnel junctions \cite{mooij,dmitriev}.
The effect of nonequilibrium Andreev states on the Josephson current-phase relation and $I_c$ in superconducting weak links and hybrid nanostructures 
has recently attracted much interest \cite{belzig,cuevas,glazman}. 

At low temperatures $T\ll T_c$ and frequencies $\omega\ll\Delta$,  the small density of quasiparticles affects neither $T_c$ nor the dynamics of superconducting condensate, yet the effects of oscillating superflow and nonequilibrium quasiparticle states on dissipative kinetic coefficients cause a strong field dependence of the surface resistance $R_s(H)$. Usually $R_s$ increases with the amplitude of the radio-frequency (rf) field \cite{hein,oates}, consistent with the expected enhancement of dissipation by pairbreaking currents, electron overheating, penetration of vortices, etc.   A remarkable departure from this conventional scenario is the puzzling reduction of $R_s$ by the rf field, which has been observed on many superconductors.  For instance, $R_s$ measured on the Nb resonator cavities at 2K and 1-2 GHz typically decreases by 10-20 $\%$ at $H\simeq 20-30$ mT and then increases at higher fields \cite{cavity,ag}. Moreover, the Nb resonators alloyed with Ti \cite{cav1} or N \cite{cav2} impurities can exhibit even stronger microwave suppression of $R_s$ (by $\simeq 50-70\%$ at 2K) which extends to the fields $H\simeq 90-100$ mT at which the density of screening currents $J\simeq H/\lambda$  reaches  $\simeq 50\%$ of the pairbreaking limit $J_d\simeq H_c/\lambda$, where $H_c$ is the thermodynamic critical field, and $\lambda$ is the London penetration depth (see Fig. 1).   Reduction of $R_s$ by dc or microwave fields has also been observed on thin films \cite{sridhar,film,delft}. The behavior of $\sigma_1(H)$ at $T\ll T_c$  is related to the fundamental limits of dissipation which controls decoherence in Josephson qubits \cite{qubits} or performance of resonator cavities for particle accelerators \cite{cavity} or microresonators \cite{caltech}.  

\begin{figure}[tb]
\includegraphics[width=6cm]{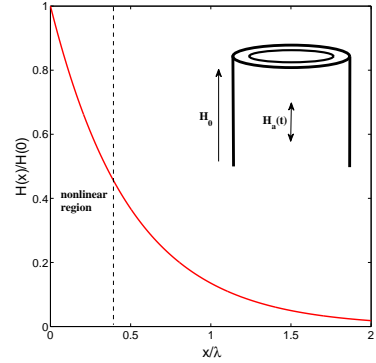}
\caption{\label{fig:fig1} Penetration of a parallel rf field into a superconductor. The dashed line 
depicts a layer where the current pairbreaking is essential. Inset shows a thin film $(d<\lambda)$ deposited onto a cylindrical substrate in a superimposed 
dc and rf field.}
\end{figure}

In this work a theory of nonlinear conductivity and the microwave suppression of $R_s$ in dirty s-wave superconductors is proposed.  
Here the electromagnetic response at weak fields is described by the local Ohmic relation $\mathbf{J}(\textbf{r},\omega)=[\sigma_1(\omega)-i\sigma_2(\omega)]\mathbf{E}(\textbf{r},\omega)$, where $\sigma_2=1/\mu_0\lambda^2\omega$, and $\sigma_1$ is the quasiparticle conductivity \cite{mb}
\begin{equation}
\sigma_1=(2\sigma_n\Delta/T)\ln(CT/\omega)e^{-\Delta/T}, \qquad T\ll T_c,
\label{mb}
\end{equation}
where $\sigma_n$ is the normal state conductivity, $C=4e^{-\gamma}\approx 9/4$, and $\gamma=0.577$. 
The logarithmic term in Eq. (\ref{mb}) comes from the convolution of the BCS density of states $\sigma_1\propto \int_\Delta^\infty N(\epsilon)N(\epsilon+\omega)e^{-\epsilon/T}d\epsilon$, 
which diverges at $\omega=0$ and $N(\epsilon)=N_0\epsilon(\epsilon^2-\Delta^2)^{-1/2}$ \cite{mb}, so smearing the gap singularities in $N(\epsilon)$  
decreases $\sigma_1$ at $\omega\ll T$. 
  
The broadening of the gap peaks in $N(\epsilon)$ and the reduction of a quasiparticle gap $\epsilon_g$ can be caused by current \cite{Bardeen} or by magnetic 
impurities \cite{magimp} which break the time reversal symmetry of  pairing electrons.  Particularly, the effect of dc current on $N(\epsilon)$ shown in Fig. 2  
was observed by tunneling spectroscopy \cite{denscurr}, in full agreement with the theory \cite{maki}.  
Under strong rf current, $N(\epsilon,t)$ oscillates between two solid curves in Fig. 2, so the peak in $\langle N(\epsilon)\rangle $ averaged over the rf period is smeared out within the energy region $\epsilon_g < \epsilon \lesssim \Delta$ of width  $\delta\epsilon = \Delta-\epsilon_g\sim (J/J_d) ^{4/3}\Delta$ at $J\ll J_d$ \cite{denscurr,maki}. This picture gives insight into one of  mechanisms of microwave reduction of $\sigma_1$: as the current-induced width     
$\delta\epsilon$ exceeds $\omega$, the energy cutoff in the logarithmic term in Eq. (\ref{mb}) changes from $\omega$ to $\delta\epsilon$. Hence, $\sigma_1\propto\ln[(J_d/J)^{4/3}T/T_c]$ decreases with $J$ if $J>(\omega/\Delta)^{3/4}J_d$ and $\omega\ll T$, so that the decrease of $\epsilon_g$ in the Boltzmann factor $e^{-\epsilon_g/T}$ has a smaller effect on $\sigma_1(H)$ at $J < (T/\Delta)^{3/4}J_d$ since $\delta\epsilon < T$. For instance, $\omega/T\sim 2\cdot 10^{-2}$ at 1 GHz at 2K. 

\begin{figure}[tb]
\includegraphics[width=6.4cm]{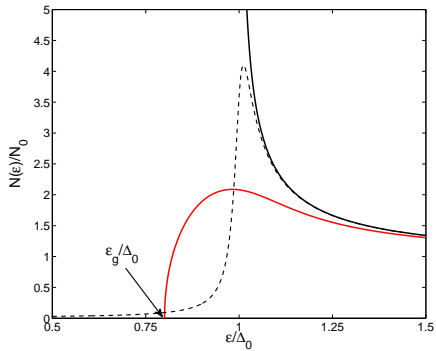}
\caption{\label{fig:fig2} The effect of current on the density of states calculated from Eqs. (\ref{card})-(\ref{v}) at $s=0.2$. The dashed line shows $N(\epsilon)=N_0\mbox{Re}[(\epsilon-i\gamma)/\sqrt{(\epsilon-i\gamma)^2-\Delta_0^2}]$ at $\gamma=0.02\Delta_0$.}
\end{figure}

A theory of $\sigma_1(H)$ must address both the pairbreaking and nonequilibrium effects caused by microwaves. 
Most of the previous works have focused on nonequilibrium states caused by absorption of photons by quasiparticles while neglecting the effect of rf superflow on $N(\epsilon)$ at weak fields $H\ll (\omega/\Delta)^{3/4}H_c$ and $\omega\gtrsim T$ \cite{mooij,dmitriev}. Here $\sigma_1(H)$ can be described by the linear response theory \cite{mb} but with a nonequilibrium quasiparticle distribution function $f(\epsilon,H)$ calculated from a kinetic equation.  Using this approach, it was shown recently that $\sigma_1(H)$ can decrease with $H_a$ as the quasiparticle population spreads to higher energies $\epsilon \gtrsim T$ \cite{delft},  similar to the mechanism of stimulated superconductivity \cite{ivlev}. This result was used to explain the reduction of $\sigma_1$ with $H_a$ observed on Al films at 5.3GHz at $350$ mK \cite{delft}. Here I consider a fundamentally different mechanism of microwave suppression of $\sigma_1(H)$ at strong, low-frequency fields with $\omega\ll T$ and $H > (\omega/\Delta)^{3/4}H_c$ for which the effect is due to the time-dependent $N(\epsilon,t)$ and a nonequilibrium distribution function controlled by oscillating superflow. In this case the Mattis-Bardeen theory is no longer applicable and $\sigma_1(H)$ is to be rederived using the Keldysh technique of nonequilibrium Green functions \cite{kopnin}. It is what was done in this work where the nonlinear conductivity $\sigma_1(H)$ was calculated for two cases: 1. A weak ac field superimposed onto the dc field  $H(t)=H_0+H_a\cos\omega t$, where the dc superflow can be used to tune $\sigma_1(H_0)$; 2. Parallel rf field $H(t)=H_a\cos\omega t$, as shown in Fig. 1. 
 
 In a type-II superconductor $(\lambda\gg\xi)$ considered here the rf field with $\omega\ll T$ does not generate new quasiparticles while $H(x,t)$ varies slowly over the coherence length $\xi$. In this case the dependence of $\mathbf{J}(\mathbf{r},t)$ on the vector potential $\mathbf{A}(\mathbf{r},t)$ is local but nonlinear and time dispersive. It can be expressed in terms of nonequilibrium matrix Green functions $\check{G}(t,t', {\bf r})$ which satisfy the time-dependent Usadel equation coupled with kinetic equations taking into account scattering of quasiparticles on phonons \cite{lo,kopnin,belzig}.  The nonlinear conductivity $\sigma_1=2\langle {\bf JE}\rangle/E_a^2$ is calculated in the Supplemental Material \cite{suppl} by averaging the dissipated power over the rf period of slowly oscillating superflow at $\omega\ll T$ and $(H/H_c)^2\ll 1$. Here ${\bf E} = - \partial_t {\bf A}= {\bf E}_a\sin\omega t$ is the electric field,  $\check{G}[\epsilon,Q(t)]$ depends on the local current density $\mathbf{J}({\bf r},t)=-\phi_0\mathbf{Q}({\bf r},t)/2\pi\mu_0\lambda^2$, where $\mathbf{Q}=\nabla\chi+2\pi\mathbf{A}/\phi_0$, $\phi_0$ is the flux quantum, $\chi$ is the 
phase of the order parameter, $\Delta(y)=\Delta e^{iQ(t)y}$. The normal and anomalous Green functions are parametrized by 
$G^R=\cosh(u+iv)$ and $F^R=e^{iQy}\sinh(u+iv)$,  where $u$ and $v$ satisfy the quasistatic Usadel equation \cite{kopnin,belzig}:
\begin{eqnarray}
\epsilon +is\cosh (u+iv)=\Delta \coth (u+iv),
\label{status} \\
s(t)=DQ^2/2=e^{-2x/\lambda}\beta(t)\Delta_0.  
\label{s}
\end{eqnarray}
Here $\beta(x,t)=(H/2H_c)^2=(J/2J_d)^2\ll 1$, $D$ is the electron diffusivity, 
$H_c=\phi_0/2^{3/2}\pi\mu_0\lambda\xi$,  $\xi=(D/\Delta_0)^{1/2}$, and  
$\Delta_0=\Delta(T=0,Q=0)$.  A correction to $Q$ due to the nonlinear Meissner effect \cite{nme} is disregarded.

The rf conductivity for the weak rf field superimposed onto the dc field is given by \cite{suppl}
\begin{gather}
\sigma_1(H_0)=\frac{2\sigma_n}{\omega}\left(1-e^{-\omega /T}\right)\int_{\epsilon_g
}^{\infty }e^{-\epsilon /T} M(\epsilon, \omega,s)d\epsilon , \label{sigdc}\\
M(\epsilon,\omega,s)=\cos v_{\epsilon}\cos v_{\epsilon +\omega }\cosh( u_{\epsilon }+ u_{\epsilon+\omega }), 
\label{M}
\end{gather}
where the spectral function $M(\epsilon,\omega,s)$ incorporates the effect of dc superflow on $N(\epsilon,Q)$ and the coherence factors. 
Here $u_\epsilon$ and $v_\epsilon$ are defined by the real and imaginary parts of Eq. (\ref{status}) which yields the cubic equation 
$\sinh ^{3}2u+[(\epsilon ^{2}-\Delta ^{2})/s^{2}+1]\sinh 2u-2\epsilon \Delta/s^{2}=0$ with the following Cardano solution:
\begin{gather}
\sinh 2u =[(r+\epsilon \Delta s) ^{1/3}-(r-\epsilon \Delta s)^{1/3}]/s, 
\label{card}\\
r=[\epsilon ^{2}\Delta ^{2}s^{2}+
(\epsilon ^{2}+s^{2}-\Delta ^{2})^{3}/27]^{1/2},
\label{r} \\
\sin v=\bigl[-\Delta +(\Delta ^{2}-s^{2}\sinh ^{2}2u)^{1/2}\bigr]/2s\cosh u
\label{v}
\end{gather}
The quasiparticle density of states and the gap energy $\epsilon _{g}$ at which 
$N(\epsilon)$ vanishes (see Fig. 2), are given by $N(\epsilon)=N_0\cosh u\cos v$, and \cite{maki}:
\begin{eqnarray}
\epsilon _{g}^{2/3}=\Delta ^{2/3}-s^{2/3},
\qquad
\Delta =\Delta_0-\pi s/4,
\label{gap}
\end{eqnarray}
where $\Delta$ is obtained from the BCS gap equation at $T=0$ in the first order in $s$ \cite{suppl}. Here $\epsilon_g(H_0)$ 
decreases with $H_0$ but remains finite $(\epsilon_g\simeq 0.3\Delta_0)$ even at the maximum superheating field $H_{sh}\approx 0.84H_c$ 
for the Meissner state \cite{sh}.
\begin{figure}[tb]
\includegraphics[width=6.3cm]{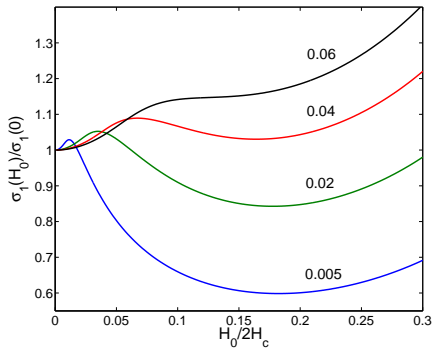}
\caption{\label{fig:fig3} Linear conductivity $\sigma_1(H_0)$ calculated from Eqs. (\ref{sigdc})-(\ref{gap}) for $\omega/\Delta_0$: $0.005$, $0.02$, $0.04$, $0.06$, and $T/\Delta_0 =0.1$. }
\end{figure}
Shown in Fig. 3 is the linear rf conductivity $\sigma_1(H_0)$ biassed by a dc superflow calculated from Eqs. (\ref{sigdc})-(\ref{gap}).  At 
$H_0=0$ and $H_a\ll (\omega/\Delta)^{3/4}H_c$,  Eqs. (\ref{sigdc})-(\ref{gap})  reproduce Eq. (\ref{mb}), but at higher field
$\sigma_1(H_0)$ has a minimum which becomes more pronounced as $\omega$ decreases. This behavior is due to 
interplay of the current-induced broadening of the gap peak in $N(\epsilon,s)$ and the reduction of $\epsilon_g$ shown in Fig. 2.  
As a result, $\sigma_1$ becomes dependent on $H_0$ if $H_0> \left( \omega /\Delta \right) ^{3/4}H_c$ and reaches minimum at $H_0\sim (T/T_c)^{3/4}H_c\ll H_c$.
The field region $(\omega/\Delta_0)^{3/4}H_c < H_0 < (T/\Delta_0)^{3/4}H_c$ where $\sigma_1(H_0)$ 
decreases with $H_0$ shrinks as $\omega$ increases and disappears at $\omega>T$, as shown in Fig. 3.          

Calculation of the nonlinear conductivity $\sigma_1(H_a)$ at a strong rf field $H(t)=H_a\cos\omega t$ requires taking temporal oscillations 
of $N(\epsilon,t)$ and $f(\epsilon,t)$ into account.  Here $\sigma_1(H_a)=2\langle {\bf JE}\rangle/E_a^2$ is defined 
as before by averaging the power over the rf period \cite{suppl}:
\begin{equation}
\sigma_1(H_a)=\frac{2\sigma_n}{\pi}\int_0^{\pi/\omega}\!\!\!\!dt\int_{\epsilon_g(t)}
^{\infty }\![f(\epsilon,s)-f(\epsilon+\omega,s)]Md\epsilon ,
\label{sigrf}
\end{equation}
where $M[\epsilon,\omega,s(t)]$ is given by Eq. (\ref{M}). Solving the kinetic equation for $f(\epsilon,s)$ with time-dependent 
parameters and the electron-phonon collision integral \cite{kopnin} is a very complicated problem,
so I only consider here the case of $\mbox{min}(\tau_r^{-1},\tau_s^{-1})\ll\omega\ll T$ for which
the rf period is shorter than either the recombination time $\tau_r$ and the scattering time $\tau_s$ of quasiparticles on phonons \cite{kaplan}
 \begin{equation}
\tau _r=\tau_1( T_c/T) ^{1/2}e^{\Delta /T}, \qquad \tau_s=\tau_2(T_c/T)^{7/2},
\label{tau}
\end{equation}
where $\tau_1$ and $\tau_2$ are materials constants. 
Taking $\Delta=1.9T_c$, $T_c=9.2$K,  $\tau _1\simeq 3\cdot 10^{-12}$ s and $\tau_2\simeq 8\cdot 10^{-11}$ s for Nb \cite{kaplan}, yields 
$\tau_r \sim 0.4\ \mu$s and $\tau_s\simeq 1.7\cdot 10^{-8}$ s at 2K.  The condition $\tau_s^{-1}<\omega<T$ that the quasiparticle density 
does not change during the rf period, can be satisfied in a frequency range, ($0.06 - 44$ GHz) relevant to many 
experiments \cite{cavity,caltech,qubits}.  

The distribution function $f(\epsilon,t)$ can be obtained from the 
following consideration. As $s(t)$ increases, $N(\epsilon,s)$ extends to lower energies as shown in Fig. 2, but because 
the quasiparticles do not scatter during the rf period if $\omega\tau_s\gg 1$, the probability to occupy the energy state $\epsilon$ moved from the state $\tilde{\epsilon}$ at $s=0$ 
does not change.  The relation between $\tilde{\epsilon}$ and $\epsilon$ follows from the conservation of states: 
$\int_{\epsilon_g}^\epsilon N(\epsilon,s)d\epsilon=\int_{\Delta_0}^{\tilde{\epsilon}}N(\epsilon,0)d\epsilon=N_0(\tilde{\epsilon}^2-\Delta_0^2)^{1/2}$, 
giving $\tilde{\epsilon}^2=\Delta_0^2+[\int_{\epsilon_g}^{\epsilon}\cosh u\cos v d\epsilon]^2$. The function $f(\tilde{\epsilon})$ 
ensures that the quasiparticle density $n_{qp}=\int f(\tilde{\epsilon})N(\epsilon,s)d\epsilon=\int_0^\infty f[\tilde{\epsilon}(\psi)]d\psi$ does not change during the rf period, 
where $\psi=\int_{\epsilon_g}^\epsilon N(\epsilon,s) d\epsilon$.  Then the condition $f(\epsilon,s)=\exp(-\tilde{\epsilon}/T)$ at $s(t)=0$ yields 
\begin{equation}
f=\exp\bigl[ -\frac{1}{T}\bigl(\Delta_0^2+\bigl[\int_{\epsilon_g}^\epsilon\cos v_\epsilon\cosh u_\epsilon  d\epsilon\bigr]^2\bigr)^{1/2}\bigr]
\label{f}
\end{equation}
 The quasiparticle temperature $T$ at $\omega\tau_s\gg 1$ is defined by the stationary power balance, $R_sH_a^{2}/2=h(T_i-T_0)=Y(T-T_i)$. Here $T_i$ is the lattice temperature, $T_0$ is the ambient temperature, $h=\kappa h_K/(dh_K + \kappa)$ accounts for heat transfer due to thermal conductivity $\kappa$ and the Kapitza interface conductance $h_K$ across a film of thickness $d$, and $Y(T)$ quantifies the energy transfer rate from quasiparticle to phonons \cite{t}. For weak overheating, the heat transfer may be linearized in $T-T_0 \ll T_0$: 
 \begin{gather}
T-T_0=\frac{\alpha T_0}{R_{s0}}\left(\frac{H_a}{H_c}\right)^2R_s(H_a,T), \label{T}\\
\alpha=\frac{R_{s0}B_c^2}{2\mu_0^2T_0}\left(\frac{1}{Y}+\frac{d}{\kappa} +\frac{1}{h_K}\right),
\label{al}
\end{gather}
where $Y$, $h$ and $R_{s0} = R_s(T_0)$ are taken at $T=T_0$ and $H_a=0$. The surface resistance $R_s(T,H_a)$ is calculated by integrating the local power
$R_sH_a^2/2=(\mu_0\omega\lambda H_a)^2\int_0^\infty e^{-2x/\lambda}\sigma_1(\beta)dx$. Changing here to integration over  
$\beta=\beta_0e^{-2x/\lambda}$ defined by Eq. (\ref{s}) yields    
\begin{equation}
R_s=\frac{\mu_0^2\omega^2\lambda^3}{2\beta_0}\int_0^{\beta_0}\sigma_1(\beta)d\beta.
\label{rs}
\end{equation} 
\begin{figure}[tb]
\includegraphics[width=6.2cm]{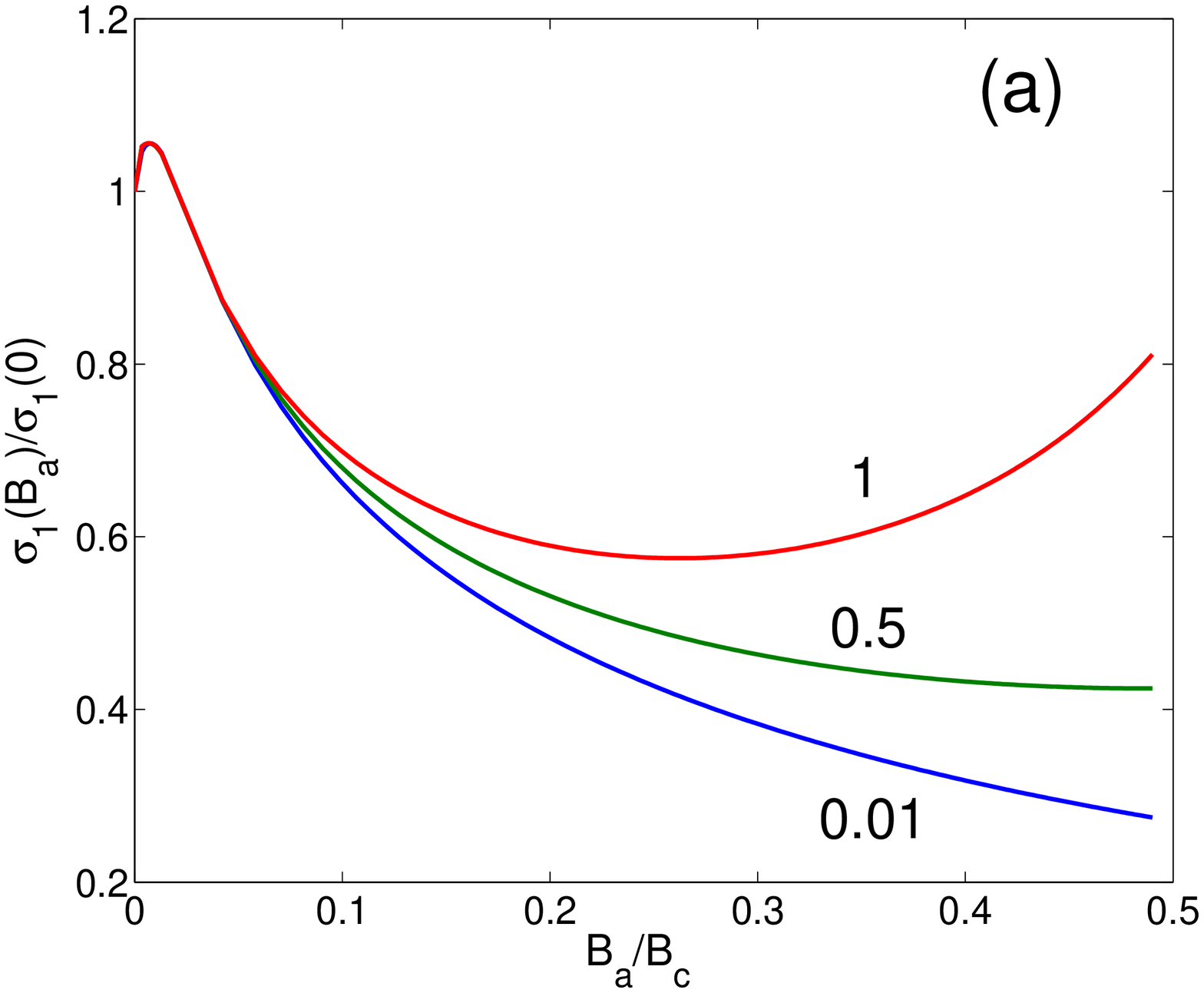}
\includegraphics[width=6.4cm]{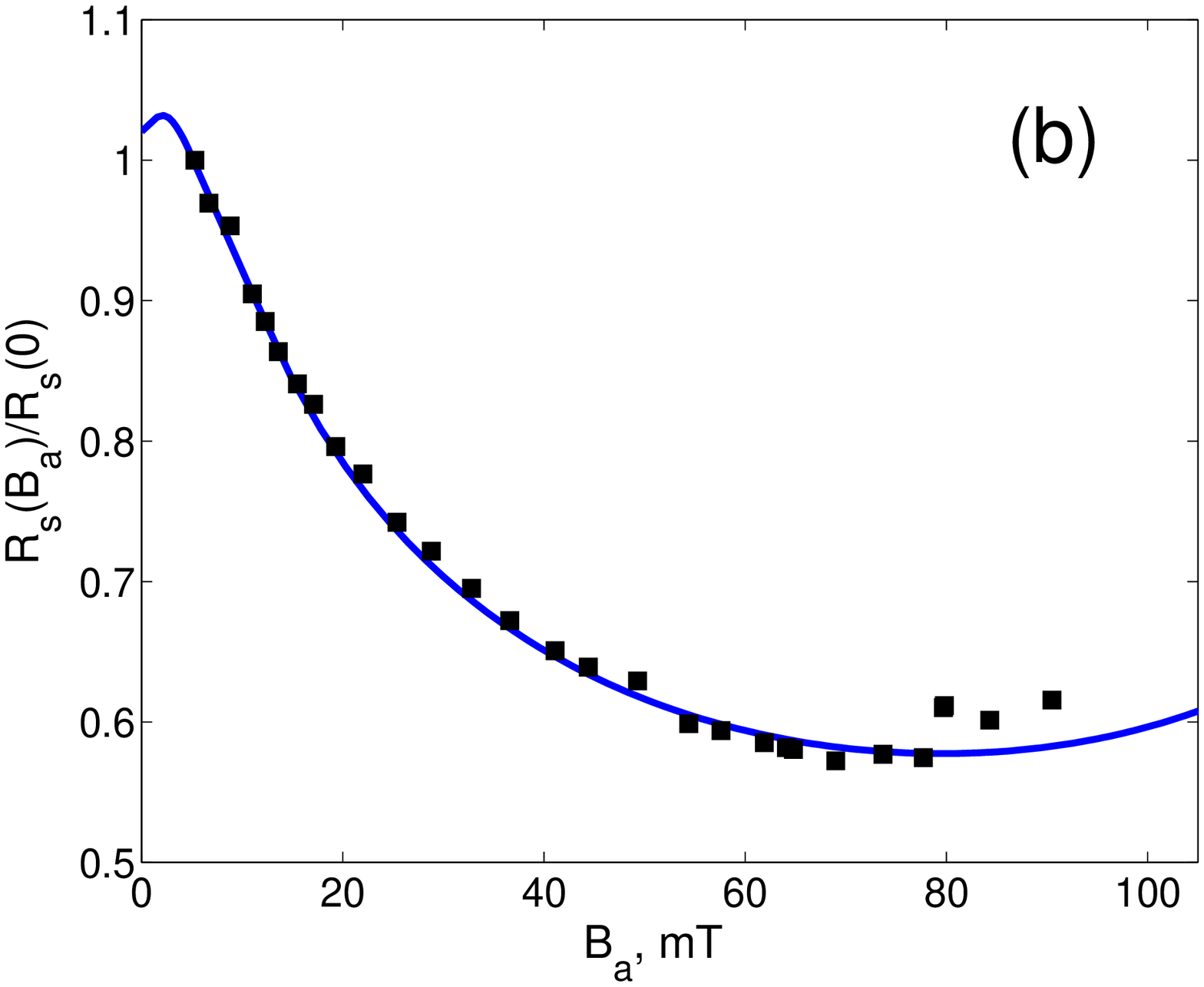}
\caption{\label{fig:fig4} (a) $\sigma_1(B_a)$ calculated from Eqs. (\ref{sigrf})-(\ref{rs}) at $T/\Delta_0 =0.1$, $\omega/\Delta_0=0.0025$ and $\alpha=$ $0.01$, $0.5$, and $1$. (b) $R_s(B_a)$ calculated for $\alpha=0.91$, $T_0=2$ K, $\Delta_0 = 17.5$ K and $B_c=200$ mT. The dots show the experimental data for the Nb cavity at 1.75 GHz \cite{cav1}. The quench at $B_a\approx90$ mT $\approx 0.5B_c$ is likely due to penetration of vortices at surface defects.}
\end{figure}
Equations (\ref{sigrf}) and (\ref{f})-(\ref{rs}) determine self-consistently $R_s(H_a,T)$ and $T(H_a)$.  
Shown in Fig. 4a is $\sigma_1(H_a)$ calculated from Eqs. (\ref{sigrf})-(\ref{al}) for different values of $\alpha$. 
The field dependence of $\sigma_1(H_a)$ is similar to that of $\sigma_1(H_0)$ in Fig. 3: in both cases the 
current-induced smearing of $N(\epsilon)$ reduces $\sigma_1(H)$, but the mechanisms of the increase of $\sigma_1(H)$ at higher fields are different.  
For a weak rf field superimposed onto the dc field, the increase of $\sigma_1(H_0)$ results from the 
reduction of $\epsilon_g(H)$, while the increase of $\sigma_1(H_a)$  in Fig. 4a is due to overheating: 
the condition that the density of quasiparticles does not change during the rf cycle greatly enhances the microwave reduction 
of $\sigma_1(H_a)$.  If $\omega \ll [\tau_s^{-1}(T),\tau_r^{-1}(T)]$ the minimum in $\sigma_1(H_a)$ 
is controlled by the field reduction of $\epsilon_g(H_a)$ as  $f(\epsilon)\to \exp(-\epsilon/T)$ becomes more equilibrium. 

Shown in Fig. 4b is $R_s(B_a)$  calculated from  Eqs. (\ref{sigrf}) and (\ref{f})-(\ref{rs}) to fit the experimental data of Ref. \cite{cav1} 
with only one adjustable parameter $\alpha=0.91$ for which the overheating $T-T_0\approx 0.17$ K calculated from Eq. (\ref{T}) is indeed weak 
even at $B_a=80$ mT,  $R_s(B_a)/R_{s0}=0.6$ and $T_0=2$K. This theory describes 
well the microwave suppression of $R_s$ observed on Ti-alloyed Nb cavities \cite{cav1}.
For $R_s=20$ n$\Omega$, $\kappa=10$ W/mK, $h_K=5$ kW/m$^2$K at 2K and $d=3$ mm \cite{cav1}, 
the phonon heat transfer in Eq. (\ref{al}) can only account for $\alpha\approx 0.06$. The larger value of 
$\alpha=0.91$ used to fit $R_s(H_a)$ in Fig. 4b indicates a significant role of electron overheating \cite{suppl,t}. 

The parameters $Y(T)$, $\tau_s$ and $\tau_r$ are not only controlled by the  
scattering and recombination of quasiparticles \cite{kaplan,t}, but also by the smearing of the gap peak in $N(\epsilon)$  
due to inhomogeneities, inelastic scattering or impurities \cite{subgap,magimp}, which can make $Y$ very sample dependent. 
Interplay of the subgap states and current pairbreaking can bring about competing mechanisms of nonlinearity of $R_s(H_a)$, since 
the subgap states can cause both a finite $\tau_r$ at $T\to 0$ \cite{reltime_klap} and a residual 
conductivity \cite{disorder_klap}. In any case, the microwave suppression of $R(H_a)$ is more pronounced for sharper gap peaks in $N(\epsilon)$ at $H_a=0$, 
so that $\tau_s$ and $\tau_r$ are not much reduced and the current-induced broadening of $N(\epsilon)$ takes over at comparatively low fields (see Fig. 2). 
This conclusion is consistent with the observed variability of the field-induced reduction of $R_s(H_a)$ \cite{cavity,ag,cav1,cav2} and the tunneling 
measurements \cite{cav1} which revealed fewer subgap states in $N(\epsilon)$ for the Nb resonators exhibiting the significant minimum in  
$R_s(H_a)$ shown in Fig. 4b.  A dc field applied parallel to a thin film can be used to  tune $\sigma_1(H_0)$ and separate 
current pairbreaking from nonequilibrium effects \cite{sridhar,Groll}.

In conclusion, a theory of nonlinear conductivity of dirty superconductors at low temperatures and strong rf electromagnetic field is developed. 
The theory explains the effect of the field-induced suppression of surface resistance,  in excellent agreement with recent experiments. 
         
This work was supported by DOE HEP under Grant No. DE-SC0010081.

\clearpage

\section{Supplemental Material.}

Calculations of $\sigma_1(H)$ were done using 
the time-dependent Usadel equations for the $4\times 4$ quasiclassical Greens function $\check{G}(\textbf{r},t,t')$ \cite{lo,kopnin,belzig}:
\begin{gather}
\partial _{t}\hat{\sigma}_z\check{G}+\partial _{t'}\check{G}\hat{\sigma}_z=D\check{\Pi}\cdot (\check{G}\cdot 
\check{\Pi}\check{G})-[\hat{\Delta}%
,\check{G}]  \label{usad} \\
\check{G}=\left( 
\begin{array}{cc}
\hat{G}^{R} & \hat{G}^{K} \\ 
0 & \hat{G}^{A}
\end{array}
\right) ,\qquad 
\hat{\Delta}=\left( 
\begin{array}{cc}
 0 & \Delta \\ 
\Delta^* & 0
\end{array}
\right),
\label{def}
\end{gather}
where $\hat{G}^{R}$ and $\hat{G}^{A}$ are the retarded and advanced Green functions, 
$\hat{G}^{K}=\hat{G}^{R}\cdot\hat{f}-\hat{f}\cdot \hat{G}^{R}$ is the Keldysh function expressed in terms of a distribution 
function of quasiparticles, $\hat{f}(\epsilon,t)$, the hat denotes matrices in the Nambu space, 
$\hat{\Pi}=\nabla +i\pi \mathbf{A}\hat{\sigma}_z/\phi _{0}$, $\check{G}\cdot \check{G}=\check{1}$, the 
dot product means time convolution,  $D$ is the diffusion coefficient, $\phi_0$ is the flux quantum, and $\hat\sigma_z$ is the Pauli matrix. 
For a dirty type-II superconductor with $\lambda\gg\xi$,  the relation between the
current density and the vector potential is local but nonlinear and time-dispersive:
\begin{gather}
{\bf J}({\bf r},t)=\frac{\sigma_n}{2}\mbox{Im}\int D(t,t',\mathbf{r}){\bf A}({\bf r},t')dt', \\
D=\mbox{Tr}\!\!\int\{\hat{G}_{z}^{R}(t,t^{\prime })\bigl[\hat{G}_{z}^{R}(t^{\prime },t_{1})f(t_{1},t) 
-\hat{G}_{z}^{A}(t_{1},t)f(t^{\prime },t_{1})\bigr] \nonumber \\
+\bigl[f(t_{1},t^{\prime})\hat{G}_{z}^{R}(t,t_{1})
-f(t,t_{1})\hat{G}_{z}^{A}(t_{1},t^{\prime })\bigr]\hat{G}_{z}^{A}(t^{\prime },t)\}dt_1
\end{gather}
Here the gradient terms $\nabla G$ were neglected, and the matrix $\hat{f}$ reduces to a single distribution function since the tranverse 
electromagnetic field does not cause the electron-hole imbalance.  

Current density was calculated using the mixed Wigner-Fourier representation 
\begin{equation}
G(t,t^{\prime})=\int_{-\infty}^{\infty}G\left(\epsilon, t_{0}\right)
e^{i\epsilon(t^{\prime}-t)}\frac{d\epsilon}{2\pi}, \quad t_0=\frac{1}{2}(t+t'),
\end{equation}
Expansion of the time convolution in small derivatives over the slow variable $t_0$ yields
\begin{gather}
\int G(t^{\prime},t_{1})f(t_{1},t)dt_{1}= 
\int e^{i\epsilon(t-t^{\prime}%
)}[  G(\epsilon,t_{0})f(\epsilon,t_{0})\nonumber \\
+\frac{i}{2}\left(  \dot
{f}(\epsilon,t_{0})\partial_{\epsilon}G(\epsilon,t_{0})-\dot{G}(\epsilon,t_{0})\partial_{\epsilon}f(\epsilon,t_{0})\right)]  \frac
{d\epsilon}{2\pi}%
\end{gather}
where the overdot means derivative with respect to $t_{0}$ \cite{lo,belzig,kopnin}. Neglecting linear in $\omega$ terms then reduces $D(t,t^{\prime})$ to:%
\begin{gather}
D =\mbox{Tr}\!\!\int\bigl[ e^{i(\epsilon-\epsilon^{\prime}%
)(t-t^{\prime})}\hat{G}_{z}^{R}(\epsilon^{\prime},t_{0})+e^{i(\epsilon
^{\prime}-\epsilon)(t-t^{\prime})}\hat{G}_{z}^{A}(\epsilon^{\prime}%
,t_{0})\bigr]  \nonumber \\
 \times\bigl[  \hat{G}_{z}^{R}(\epsilon,t_{0})-\hat{G}_{z}^{A}(\epsilon
,t_{0})\bigr] f(\epsilon,t_{0})\frac{d\epsilon d\epsilon^{\prime}}%
{(2\pi)^{2}}%
\end{gather}

The time-averaged nonlinear conductivity $\sigma_1$ is defined in terms of the mean dissipated power 
$q=\sigma_1E_0^2/2$ induced by the ac electric field $\mathbf{E}=-\partial_{t}\mathbf{A=E}_{0}\sin\omega t$ for which
$\mathbf{A}(t)=(\mathbf{E}_{0}/\omega)\cos\omega t$. Here
\begin{gather}
q=\lim_{t_{m}\rightarrow\infty}\frac{1}{2t_{m}}\int_{-t_{m}}^{t_{m}}%
\mathbf{J}(t)\mathbf{E}(t)dt=  \\
\frac{i\sigma_{n}E_{0}^{2}}{4t_{m}\omega}%
\int_{-t_{m}}^{t_{m}}\sin\omega tdt\int dt^{\prime}%
\cos\omega t^{\prime}\int\frac{d\epsilon d\epsilon^{\prime}}{(2\pi)^{2}%
}\times \nonumber \\
\bigl[ e^{i(\epsilon-\epsilon^{\prime})(t-t^{\prime})}P_{1}(\epsilon
,\epsilon^{\prime},t_{0})+e^{i(\epsilon^{\prime}-\epsilon)(t-t^{\prime})}%
P_{2}(\epsilon,\epsilon^{\prime},t_{0})\bigr] \nonumber
\end{gather}
where $P_{1}=\mbox{Tr}\hat{G}_{z}^{R}%
(\epsilon^{\prime},t_{0})[\hat{G}_{z}^{R}(\epsilon,t_{0})-\hat{G}%
_{z}^{A}(\epsilon,t_{0})]  f(\epsilon,t_{0})$ and $P_{2}=\mbox{Tr}\hat{G}_{z}^{A}(\epsilon^{\prime},t_{0})[
\hat{G}_{z}^{R}(\epsilon,t_{0})-\hat{G}_{z}^{A}(\epsilon,t_{0})]
f(\epsilon,t_{0}).$ Changing variables $t=t_{0}+t_{1}/2$ and $t^{\prime
}=t_{0}-t_{1}/2,$ yields 
\begin{gather}
q=\lim_{t_{m}\rightarrow\infty}\frac{i\sigma_{n}E_{0}^{2}}{8t_{m}\omega}%
\int_{-t_{m}}^{t_{m}} dt_{0}dt_{1}[\sin\omega t_{1}+\sin2\omega t_{0}] \times \nonumber \\
\int\frac{d\epsilon
d\epsilon^{\prime}}{(2\pi)^{2}}\bigl[  e^{i(\epsilon-\epsilon^{\prime})t_{1}%
}P_{1}(\epsilon,\epsilon^{\prime},t_{0})+e^{i(\epsilon^{\prime}-\epsilon
)t_{1}}P_{2}(\epsilon,\epsilon^{\prime},t_{0})\bigr]. 
\end{gather}
In the dirty limit $P_{1}(\epsilon,\epsilon^{\prime},t_{0})$ and
$P_{2}(\epsilon,\epsilon^{\prime},t_{0})$ are even functions of $t_{0}$, so
only $\sin\omega t_{1}$ contributes:
\begin{eqnarray}
\sigma_1  =
\frac{\sigma_{n}}{4\pi}\int_{0}^{\pi/\omega}dt\int d\epsilon[  P_{1}(\epsilon,\epsilon+\omega,t)- \nonumber \\
P_{1}(\epsilon+\omega
,\epsilon,t)+P_{2}(\epsilon+\omega,\epsilon,t)-P_{2}(\epsilon,\epsilon
+\omega,t)]
\label{sig}
\end{eqnarray}
To calculate $\mbox{Tr}\hat{G}_{z}\cdot \hat{G}_{z}=2G\cdot G+F\cdot F^{\dag}+F^{\dag}\cdot F$ in $P_1$ and $P_2$,
it is convenient to use the parameterization $\hat{G}^R(\epsilon, Q)=-\hat{G}^{A*}(\epsilon, -Q)$, where \cite{belzig}
\begin{equation} 
\hat{G}^R=\left( 
\begin{array}{cc}
\cosh(u+iv) & \sinh(u+iv)e^{iQy} \\ 
-\sinh(u+iv)e^{-iQy} & -\cosh(u+iv)
\end{array} \right),
\label{parametr}
\end{equation}
The supercurrent phase factors in $F=Fe^{iQy}$ and $F^\dagger=F^\dagger e^{-iQy}$ cancel out in $\mbox{Tr}\hat{G}_{z}\cdot \hat{G}_{z}$, 
so that
\begin{gather}
P_{1}(\epsilon,\epsilon^{\prime})-P_{2}(\epsilon,\epsilon^{\prime}%
)=2\{[G^{R}(\epsilon^{\prime})-G^{A}(\epsilon^{\prime})][G^{R}(\epsilon
)-G^{A}(\epsilon)] \nonumber \\
+[F^{R}(\epsilon^{\prime})-F^{A}(\epsilon^{\prime}%
)][F^{R}(\epsilon)-F^{A}(\epsilon)]\}f(\epsilon,t)
\end{gather}
Here $G^{R}(\epsilon)-G^{A}(\epsilon)=2\cosh u\cos v$ and $F^{R}(\epsilon)-F^{A}(\epsilon)=2\sinh u\cos v$.  
Changing integration in Eq. (\ref{sig}) to positive energies yields
\begin{gather}
\sigma_1=\frac{2\sigma_{n}}{\pi}\int_{0}^{\pi/\omega}\!\!\!dt\int_{\epsilon
_{g}}^{\infty}\!\![\cosh u_{\epsilon}\cosh u_{\epsilon+\omega}+\sinh
u_{\epsilon}\sinh u_{\epsilon+\omega}]\nonumber \\
\times \cos v_{\epsilon}\cos v_{\epsilon
+\omega}[f(\epsilon,s)-f(\epsilon+\omega,s)]d\epsilon
\end{gather}
which reduces to Eq. (4), (5) and (10) of the main text. Here $u$ and $v$ satisfy the quasi-static Usadel equation,
\begin{eqnarray}
\epsilon +is\cosh (u+iv)=\Delta \coth (u+iv),
\label{status} \\
s(t)=DQ^2/2=e^{-2x/\lambda}\beta(t)\Delta_0.  
\label{s}
\end{eqnarray}
Here $s$ and $\beta(x,t)=(H/2H_c)^2=(J/2J_d)^2\ll 1$ are the current pairbreaking parameters, 
$H_c=\phi_0/2^{3/2}\pi\mu_0\lambda\xi$,  $\xi=(D/\Delta_0)^{1/2}$, and  
$\Delta_0=\Delta(T=0,Q=0)$.  A correction to $Q$ due to the nonlinear Meissner effect is disregaded. Separation of the 
imaginary part in Eq. (\ref{status}) yields Eq. (8) in the main text which exresses $v$ in terms of $u$. The resulting qubic equation 
$$
\sinh ^{3}2u+[(\epsilon ^{2}-\Delta ^{2})/s^{2}+1]\sinh 2u-2\epsilon \Delta/s^{2}=0 
$$
has the Cardano solution (7)-(8) of the main text.

For $s=0$, we have $\cos v_{\epsilon}=1$, $\cosh
u_{\epsilon}=$ $\epsilon/\sqrt{\epsilon^{2}-\Delta^{2}}$, $\sinh u_{\epsilon
}=\Delta/\sqrt{\epsilon^{2}-\Delta^{2}}$, and $\epsilon_{g}(t)=\Delta.$ Then
$\sigma_1$ reproduces the Mattis-Bardeen result for $\omega<\Delta$:%
\begin{equation}
\sigma_1=\frac{2\sigma_{n}}{\omega}
\int_{\Delta}^{\infty}\frac{[\epsilon(\epsilon+\omega)+\Delta^{2}%
][f(\epsilon)-f(\epsilon+\omega)]d\epsilon}{\sqrt{\epsilon^{2}-\Delta^{2}%
}\sqrt{(\epsilon+\omega)^{2}-\Delta^{2}}}%
\end{equation}
At $\exp(-\Delta/T)\ll1$, we have $f(\epsilon)-f(\epsilon
+\omega)=\left(  1-e^{-\omega/T}\right)  e^{-\epsilon/T}$ so the main
contribution to this integral comes from a narrow range of energies
$\epsilon-\Delta\sim T\ll\Delta$ where $z=\epsilon-\Delta$ and
$\epsilon^{2}-\Delta^{2}\approx 2\Delta z$. Then
\begin{equation}
\sigma_1=\frac{2\sigma_{n}}{\omega}\left(  1-e^{-\omega/T}\right)
e^{-\Delta/T}\int_{0}^{\infty}\frac{e^{-z/T}dz}{\sqrt{z(z+\omega)}}%
\end{equation}
Hence  
\begin{equation}
\sigma_1=\frac{4\sigma_{n}\Delta}{\omega}\sinh\left[  \frac
{\omega}{2T}\right]  K_{0}\left[  \frac{\omega}{2T}\right]
e^{-\Delta/T},
\label{mb}
\end{equation}
where $K_0$ is the modified Bessel function. In the 
limit of $\omega \ll 2T$, Eq. (\ref{mb}) gives Eq. (1) of the main text 

The effect of current on $\Delta$ can be calculated using the gap equation in the Matsubara representation
\begin{equation}
1=2\pi T\lambda_{bcs}\sum_{\omega_n>0}^{\Omega}\frac{1}{\sqrt{(\omega_n+sg)^2+\Delta^2}},
\label{d1}
\end{equation}
where $\omega_n=\pi T(2n+1)$, $n=0,\pm 1, ...$, $\lambda_{bcs}$ is the BCS coupling constant, $\Omega$ is the Debye cutoff frequency, 
$g$ is the the normal quasilassical Green function which satisfies the thermodynamic Usadel equation $\omega_n f + sfg=\Delta g$ for uniform current flow, and $g^2+f^2=1$.   For weak currents $s\ll \Delta_0$, we can take here $g=\omega_n/\sqrt{\omega_n^2+\Delta^2}$, $\Delta = \Delta_0+\delta \Delta$, where $\delta \Delta$ is a small current-induced correction to $\Delta_0$ which satisfies the gap
 equation (\ref{d1}) at $s=0$.  Linearizing Eq. (\ref{d1}) in small $s$ and $\delta\Delta$ gives:
\begin{equation}
\sum_{\omega_n>0}\frac{s\omega_n^2}{(\omega_n^2+\Delta_0^2)^2}+\sum_{\omega_n>0}\frac{\Delta_0\delta\Delta}{(\omega_n^2+\Delta_0^2)^{3/2}}=0
\label{d2}
\end{equation} 
At low temperatures, $T\ll\Delta_0$, the sumation in Eq. (\ref{d2}) can be replaced with integration over $\omega_n$, giving $\delta\Delta = - \pi s/4$, 
which was used in Eq. (9) of the main text.
 
\subsection{Estimate of electron overheating}

A rough estimate of electron overheating can be made using the power balance, $\sigma _{1}(T)E^{2}\simeq 2T_0[S(T)-S(T_0)]\tau _r^{-1}(T)$, where $S(T)=2N_0\Delta (2\pi\Delta/T)^{1/2}\exp(-\Delta/T)$ is the entropy of quasiparticles, $\sigma_1$ is the Mattis-Bardeen conductivity and $\tau_r$ is the quasiparticle recombination time, 
and $E=\omega\lambda B_a$ is the induced electrtic field.
At $\exp(-\Delta/T)\ll 1$, all pre-exponential factors can be taken at $T=T_{0}$, so that
\begin{gather}
e^{-\Delta /kT}-e^{-\Delta /kT_{0}}=p(H_a/H_c)^2, \label{T}\\
p=\frac{\Delta_0\tau_1}{24\hbar}\left(\frac{\hbar\omega}{kT_0}\right)^2\left(\frac{\pi k T_c}{2\Delta_0}\right)^{1/2}\ln\frac{9kT_0}{4\hbar\omega}.
\label{al}
\end{gather}
Here  the relations $\xi^2=\ell\xi_0$, $\xi_0=\hbar v_F/\pi\Delta$, and $N_0/\sigma_n=3/2e^2v_F\ell$ were used,  and the Boltzmann and the Plank constants $k$ and $\hbar$ were restored. 
Linearizing Eqs. (\ref{T}) in $T-T_0$, reduces Eqs. (\ref{T}) and (\ref{al}) to Eqs. (13) of the main text with
\begin{equation}
\alpha =\frac{kT_0\tau_1}{24\hbar}\left(\frac{\hbar\omega}{kT_0}\right)^2\left(\frac{\pi k T_c}{2\Delta_0}\right)^{1/2}e^{\Delta_0/kT_0}\ln\frac{9kT_0}{4\hbar\omega}
\label{alph}
\end{equation} 
For the numbers used in the main text, $\Delta=17.5$ K and  $\tau _1\simeq 3\cdot 10^{-12}$ s for Nb at $T_0=2$ K and 1.75 GHz, Eq. (\ref{alph}) yields $\alpha \simeq 1.32$,
of the same order of magnitude as $\alpha=0.91$ used to fit the experimental data in Fig. 4b. Taking into account another cooling channel due to emission of phonons by quasiparticles $(\tau_s)$ reduces the overheating parameter $\alpha$ which can also be rather sample-sensitive due to the broadening of the peaks in $N(\epsilon)$ by subgap states and uncertainities in 
materials and superconducting parameters affecting $Y$.

\end{document}